\documentclass[runningheads]{svmult}
\usepackage{makeidx}   
\usepackage{graphicx}  
\usepackage{epsfig}
\usepackage{subeqnar}  
\usepackage{multicol}  
\usepackage{cropmark} 
\usepackage{physprbb}  
\newcommand{\mincir}{\raise -2.truept\hbox{\rlap{\hbox{$\sim$}}\raise5.truept
\hbox{$<$}\ }}
\newcommand{\magcir}{\raise -2.truept\hbox{\rlap{\hbox{$\sim$}}\raise5.truept
\hbox{$>$}\ }}
\def\hmpc{\ifmmode\,h^{-1}\,{\rm Mpc}\;\else$h^{-1}\, {\rm Mpc}\;$\fi}
\def\kmpc{\ifmmode\,h\,{\rm Mpc^{-1}}\;\else$h\,$Mpc^{-1}\;\fi}

\begin{document}
\title*{Large--Scale Structure from Galaxy and Cluster Surveys
}
\toctitle{Large--Scale Structure from Galaxy 
\protect\newline and Cluster Surveys}
%
%
\titlerunning{Large--Scale Structure}
%
\author{Luigi Guzzo
}
%
\authorrunning{Luigi Guzzo}
%
%
\institute{INAF - Osservatorio Astronomico di Brera, Via Bianchi 46,
I-23807 Merate (LC), Italy }

\maketitle              

\begin{abstract}

I review the status of large-scale structure studies based on redshift
surveys of galaxies and clusters of galaxies.  In particular, I
compare recent results on the power spectrum and two-point correlation
correlation function from the 2dF and REFLEX surveys, highlighting the
advantage of X-ray clusters in the comparison to cosmological models,
given their easy-to-understand mass selection function.  Unlike for
galaxies, this allows the overall normalization of the power spectrum
to be measured directly from the data, providing an extra constraint
on the models.  In the context of CDM models, both the shape and
amplitude of the REFLEX P(k) require, consistently, a low value for
the mean matter density $\Omega_M$.  This shape is virtually
indistinguishable from that of the galaxy power spectrum measured by
the 2dF survey, simply multiplied by a constant cluster-galaxy bias
factor.  This consistency is remarkable for data sets which use
different tracers and are very different in terms of selection
function and observational biases.  Similarly, the knowledge of the
power spectrum normalization yields naturally a value $b\simeq 1$ for
the bias parameter of $b_J$-selected (as in 2dF) galaxies, also in
agreement with independent estimates using higher-order clustering
and CMB data.  In the final part, I briefly describe the measurements
of the matter density parameter from redshift space distortions in
galaxy surveys, and show evidence for similar streaming motions of
clusters in the REFLEX redshift-space correlation function
$\xi(r_p,\pi)$.  With no exception, this wealth of independent
clustering measurements point in a remarkably consistent way towards a
low-density CDM Universe with $\Omega_M\simeq 0.3$.

\end{abstract}

\section{Introduction}
The last couple of years have witnessed an impressive series of
achievements in the field of large-scale structure, thanks to
\footnote{Review to appear in {\it DARK2002, 4th Heidelberg
Int. Conf. on Dark Matter in Astro- and Particle Physics}, (Cape Town,
February 2002), H.-V. Klapdor-Kleingrothaus \& R. Viollier eds., Springer}
new large surveys of galaxies and clusters of galaxies. The enthusiasm
for new results on the clustering of galaxies and clusters has been
strenghtened by the unprecedented possibility to couple these to the
anisotropies in the cosmic microwave background over an overlapping
range of scales (see contributions by Melchiorri and Cooray, this
volume).

In this brief review I have tried and provide a general guide for the
non-specialist through some of the large-scale structure results.
Clearly, such a review is far from being complete, although the
references indicated should allow the reader to find further links to
the available literature on the subject (before February 2002).  I
therefore apologize to those colleagues whose work has not been
adequately covered.

\section{Cosmological Background}


The currently popular model for the origin and evolution of
structure in the expanding Universe is the Cold Dark Matter (CDM)
model \cite{CDM}, whose global features provide a framework which is
remarkably consistent with a large number of observations.  The
``Cosmology 2000'' version of the model (often referred to in the
recent literature as the
'concordance' model), which takes into account the independent
evidences for a flat geometry (from the angular power spectrum of
anisotropies in the Cosmic Microwave Background \cite{CMB}) and an
accelerated expansion (from the luminosity-distance relation of
distant supernovae, used as ``standard candles'' \cite{SNIa}) is one
where CDM, in the form of some kind of weakly-interacting
non-relativistic particles (see pertinent articles in this volume),
contributes about 25-30\% of the total density, with the remaining 70\%
provided by a ``dark energy'' associated to a {\it Cosmological
Constant}.  I will comment at the end of this review on how
comfortable we should feel in front of the number of ``unseen''
ingredients of this model.  Here we shall use the model as it is, in
fact ``just a model'', i.e. a physically motivated machinery which
works remarkably well when confronted with a variety of observations.

\begin{figure}
\centering
\epsfxsize=12cm 
\epsfbox{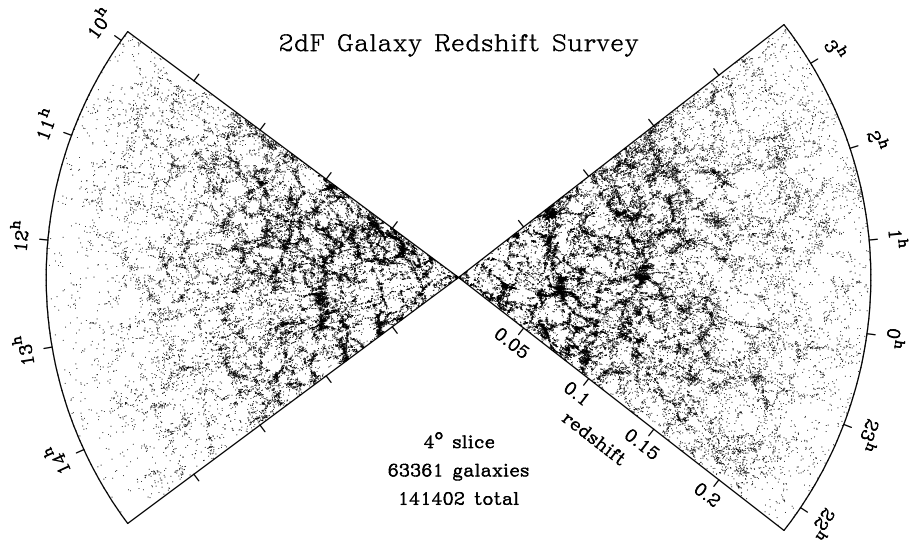} 
\caption{The distribution of over 63,000 galaxies in two 4-degree thick
slices extracted from the total of more than 210,000 galaxies that
currently make up the 2dF Galaxy Redshift Survey (2dFGRS, figure from
\cite{2dF_xipz}).
}
\label{2dF_cones}
\end{figure}
Choosing CDM (or any other model) means specifying a {\it Transfer
Function} $T(k)$.  This can be thought of as describing a linear
amplifier\footnote{$k$ is the Fourier 
wavenumber, i.e. the inverse of a 3D spatial scale $\lambda=2\pi/k$,
measured in $\hmpc$, with $h$ being the Hubble constant in units of
$100 \hmpc$.  Most recent determinations indicate $h\simeq 0.7$ with
about 10\% error, see W. Freedman contribution to this volume.} which
filters the primordial spectrum of fluctuations (typically of the
scale-invariant form $P_o(k)\propto k$ generally predicted by
inflation) to produce the shape of power spectrum we can still observe
today on large [$k\mincir 2\pi/(10 \hmpc) $] scales, $P(k) = |T(k)|^2
P_o(k)$ \cite{JAP,Paddy}.  One of the nice features of the CDM spectral
shape in any of 
its variants is to naturally lead to a {\sl hierarchical} growth of
structures, where larger entities are continuously formed from the
assembly of smaller ones \cite{White_Rees}.  Within the {\sl gravitational instability}
picture, the formation of galaxies and larger structures is completely
driven by the gravitational field of the dark matter, with our
familiar {\sl baryonic} matter representing only a tiny bit of the
mass ($2-4 \%$ of the total energy density).  The lighting-up of
galaxies and other luminous objects depends then on how the baryons
cool within the dark matter haloes and form stars, ending up as 
the only directly visible peaks of a much larger, invisible structure.

This increasing complexity in the physics involved in this cascade of
processes is reflected by the
limits in the predicting power of current detailed models of galaxy formation.
Predictions from purely gravitational n-body experiments concerning
the overall clustering of the dark mass can be regarded as fairly robust
\cite{HubbleVolume}.  More complex semi-analytical calculations
addressing the history of galaxy formation have seen exciting progress
during the last few years \cite{Somerv,Kauf,Durh,GRASIL}, but they clearly
still depend on a large number of not fully constrained parameters.

Direct measurements of large-scale structure are a classical
test-bench for CDM models and they have, for example, been the reason
for rejecting the original Einstein-DeSitter ($\Omega_{Matter}=1$)
version of the
model, whose transfer function is inconsistent with the observed
balance of large- to small-scale power \cite{APM90}.  The main
problem in the game is that true direct measurements of mass
structure (as e.g. through peculiar velocities or gravitational
lensing cosmic shear), 
are not trivial: most observations have necessarily to
use radiating objects as tracers of the mass distribution, and thus
need to go through the uncertainties mentioned above to allow
meaningful comparison to model predictions \cite{BG2001}.

\section{Progress in Large-Scale Structure Observations}

\subsection{Galaxy Redshift Surveys}

Since the 1970's, redshift surveys of galaxies have represented the
primary way to reconstruct the 3D topology of the Universe
\cite{rood}. Year 2000 has seen the completion and public release
\cite{2dF_release} of the first 100,000 galaxy redshift measurements
by the Anglo-Australian 2dF Galaxy Redshift Survey\footnote{\tt
http://www.mso.anu.edu.au/2dFGRS}, the largest complete sample of
galaxies with measured distances to date \cite{2dF_release,JAP_texas}.  This
survey includes all galaxies with blue magnitude $b_J$ brighter than
$\sim 19.5$, mainly over two areas covering $\sim 2000$ square degrees
in total, to an effective depth of about $600 \hmpc$ ($z\sim 0.2$).
Its immediate precursors \cite{LCRS,ESP} reached a
similar depth, but over much smaller areas: for comparison, the Las
Campanas Redshift Survey (LCRS \cite{LCRS}),
measured a total of 16,000 redshifts, against the 250,000 that
will eventually form the full 2dF survey.  A plot of the galaxy
distribution within the two main sky regions of this survey is shown
in Fig.\ref{2dF_cones}.  Here one can appreciate in detail the wealth
of structures typical of the distribution of galaxies: clusters,
superclusters (filamentary or perhaps sheet-like) and {\sl voids},
i.e. regions of very low galaxy density \cite{rood}.  I will discuss
the main clustering results from this survey in the following sections.

In a parallel effort, the Sloan Digital Sky Survey (SDSS)\footnote{\tt
http://www.sdss.org/} is covering a large fraction of the Northern sky
with a uniform five-band CCD survey ($u^\prime$, $g^\prime$,
$r^\prime$, $i^\prime$, $z^\prime$), plus measuring redshifts for one
million galaxies over the same area \cite{SDSS_overview}.  The
photometry reaches a red magnitude $r^\prime \sim 23$; the redshift
survey is limited to galaxies brighter than $r^\prime = 17.7$,
resulting in a depth similar to that of ESP, LCRS and 2dF, but over a
very large area.  Early data over 462 square degrees have been
recently released\footnote{\tt http://archive.stsci.edu/sdss/}. The SDSS
represents the largest and most comprehensive galaxy survey work ever
conceived: in addition to the redshift survey, the multi-band
photometry is going to be of immense value for a number of studies, as
estimating {\sl photometric redshifts} \cite{AFSoto,Yee} to much
larger depth, or selecting samples with well-defined colour/morphology
properties.  A relevant example of such applications has been the
discovery of several high-redshift ($z>5$) quasars, including the
$z=6.28$ case for which the first possible detection of the
long-sought Gunn-Peterson effect, essentially the fingerprint of the
``dark-ages'', has been recently reported \cite{Gunn-Peterson}.
Another important application will be the selection of about $10^5$
``red luminous'' galaxies with $r^\prime<19.5$, that will be observed
spectroscopically providing a nearly volume-limited homogeneous sample
out to $z\simeq 0.5$, to study the clustering power spectrum on
extremely large scales \cite{Eisenstein_LRG}.

Both the 2dF and SDSS redshift surveys rely upon the large multiplexing
performances of fiber-fed spectrographs, that allow the light from
several hundred galaxies over a field of view of 1-2 degrees to be
conveyed into the same slit on the spectrograph.  This specific
technology, in various forms, has been the key to the explosion of the
redshift survey industry in the 1990's, bringing the efficiency from the $10$
redshifts/night for galaxies brighter than blue magnitude $b \sim 14$ of
the 1970's, to the current $2500$ redshifts/night 
to $b \sim 19.5$ (see e.g. \cite{Texas99} for a more
accurate account).

\subsection{Surveys of X-ray Clusters of Galaxies}

Clusters of galaxies are complementary tracers of large-scale
structure (see 
e.g. \cite{Bob_Sesto}).  Especially before the current era, when 
$N>100,000$ galaxy redshifts are becoming available over
comparable volumes, groups and clusters have represented the most
efficient alternative to map very large volumes of the Universe,
exploring in this way the gross structure and its statistical
properties in the weak clustering regime.   
\begin{figure}
\centering
\epsfysize=12cm 
\epsfbox{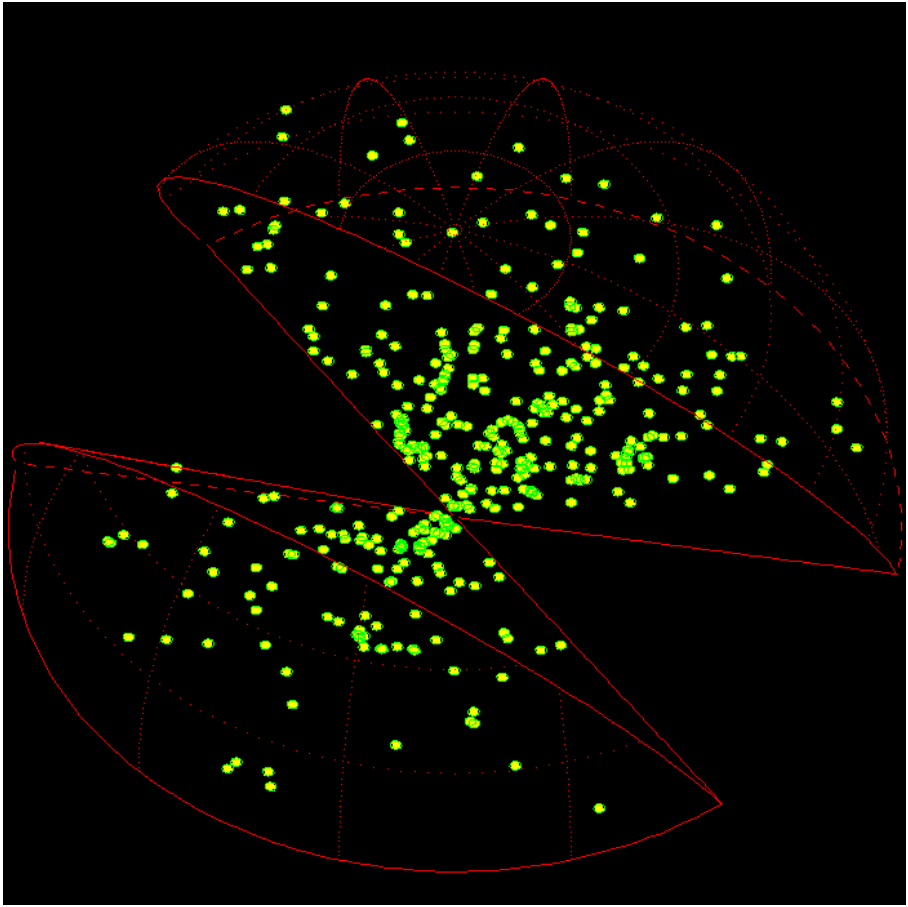} 
%
\caption{The spatial distribution of X-ray clusters in the REFLEX
survey, out to $600 \hmpc$ (from \cite{BG2001}).   Note that here
each point corresponds to a cluster, containing hundreds or thousands
of galaxies.  Structure is here mapped in a coarse way, yet sufficient
to evidence very large structures as the ``chains'' of clusters
visible in this picture. } 
\label{fig:reflex_cone}
\end{figure}

X-ray selection represents currently the most physical way to
identify large homogeneous samples of clusters of
galaxies\footnote{A notable  
powerful alternative, so far limited by technical development, is
represented by radio surveys using the 
Sunyaev-Zel'dovic effect.  In this case one measures, in the radio
domain, the CMB spectral distortions produced in the direction of a
cluster by the Inverse Compton scattering of the CMB photons over the
energetic electrons of the intracluster plasma (see
e.g. \cite{Bartlett_Sesto} for a review). } (see also discussion in
\cite{Ellis_Sesto}).  
Clusters shine in
the X-ray sky due to the {\it bremsstrahlung} emission produced by
a hot plasma ($kT\sim 1-10$ KeV) trapped within their potential
wells.  The bolometric emissivity (i.e. the energy released per
unit time and volume) of this thin gas is proportional to its density
squared and to $T^{1/2}$.
Such dependence on $n^2$ makes clusters stand out more in the
X-rays than in the optical light distribution ($\propto n$).  

Under the assumption of hydrostatic equilibrium, the intracluster gas
temperature, measured through the X-ray spectrum, is a direct probe
of the cluster mass: 
$kT\propto \mu m_p \sigma_v^2 \sim G\mu m_p M_{vir}/(3 r)$ (where
$m_p$ is the proton mass, $\mu\simeq 0.6$ the gas mean molecular
weight, $\sigma_v$ the galaxy 1D velocity dispersion and $M_{vir}$
the cluster virial mass).  
X-ray luminosity, a more directly observable quantity with current
instrumentation, shows a good correlation with temperature,
$L_{X}\propto T^\alpha$ with $\alpha\simeq 3$ and a scatter
$\mincir 30\%$.  
The practical implication,
even only on a phenomenological basis, is that clusters
selected by X-ray luminosity are in practice mass-selected, with an
error $\mincir 35$ \% (see e.g. \cite{BorganiRDCS2001} and references
therein for a more critical discussion).  Last, but not least,
the selection function of an X-ray cluster survey can be determined to
high accuracy, knowing the properties of the X-ray telescope used, in
a similar way to what is usually done with magnitude-limited samples
of galaxies \cite{Rosati_1}.  This is of fundamental importance if one
wants to compute statistical quantities and test
cosmological predictions as, e.g., the mean density or the
clustering of clusters above a given mass threshold \cite{BG2001}.

Fig.~\ref{fig:reflex_cone} plots the large-scale distribution of X-ray
clusters from the REFLEX (ROSAT-ESO Flux Limited X-ray) cluster
survey, the largest redshift survey of X-ray clusters with homogeneous
selection function to date \cite{REFLEX_survey_paper}.  This data set,
completed in 2000 and publicly released at the
beginning of 2002, is based on the X-ray all-sky survey performed
by the ROSAT satellite in the early 1990's (see
e.g. \cite{Henry,Piero_ARAA} for a comprehensive summary).  
REFLEX includes 452 clusters over the southern celestial hemisphere 
and is more than 90\% complete to a flux limit of 
$3 \times 10^{-12}$ erg s$^{-1}$ cm$^{-2}$ (in the ROSAT energy band, 0.1-2.4
keV). 
The volume explored is larger than that of the 2dF survey and
comparable to the volume that will be filled by the SDSS
1-million-galaxy redshift survey\footnote{The SDSS will however probe
a much larger volume through the luminous-red galaxy sample that will
extend to $z\sim 0.5$ \cite{Eisenstein_LRG}.}.  The fine structure
visible in Fig.~1 is obviously lost in the cluster distribution;
however, a number of cluster agglomerates and filamentary structures
with large sizes ($\sim 100 \hmpc$) are evident, showing that
inhomogeneities are still strong on such very large scales.

\section{Statistical Properties of Clustering}

\subsection{The Power Spectrum of Fluctuations}

Large-scale structure models as CDM are specified in terms of a
specific shape for the power spectrum of density fluctuations $P(k)$.
Analogously to standard signal theory, the power spectrum describes
the squared modulus of the amplitudes $\delta_k$ (at different spatial
wavelengths $\lambda=2\pi/k$) of the Fourier components of the
fluctuation field $\delta = \delta\rho/\rho$ \cite{JAP}.  Studying the
power spectrum of the distribution of luminous objects on sufficiently
large scales, where the growth of clustering is still independent of
$k$, we hope to recover a relatively undistorted information to test
the models.  

The uncertainties in relating the observed $P(k)$ of, e.g.,
galaxies to that from the theory are due to (a) nonlinear effects that
modify the linear shape below some scale; (b) the unknown relation
between the distribution of the light and that of the mass, that is what
the models predict.  
The first problem 
can be circumvented by pushing redshift surveys to larger and larger
scales, as to work well into the linear regime (and/or following
nonlinear evolution through numerical simulations).  The second one
involves knowing the {\it bias} parameter $b$.

The bias parameter can be defined
either in integral terms, through the ratio
between the variances in galaxy counts and in the mass density 
$b = \left({\delta n_{gal}(r) /
\left<n_{gal}\right>}\right)_{rms} / \left({\delta\rho(r) / 
\left<\rho\right>}\right)_{rms}$, or differentially as
$b^2=P_{gal}(k)/P_{mass}(k)$.  
For galaxy surveys, $b$ can only be deduced using additional external
constraints on the power spectrum normalization (e.g. from CMB
anisotropies, that directly probe 
fluctuations in the mass \cite{Lahav02}), or studying higher-order
moments of the distribution \cite{Verde}.  X-ray selected clusters, on
the other hand, have a specific advantage in this respect, as their bias
factor can be computed directly once the sample selection function is
known (e.g. \cite{Moscardini2000}).

\begin{figure}
\centering
\epsfxsize=12cm 
\epsfbox{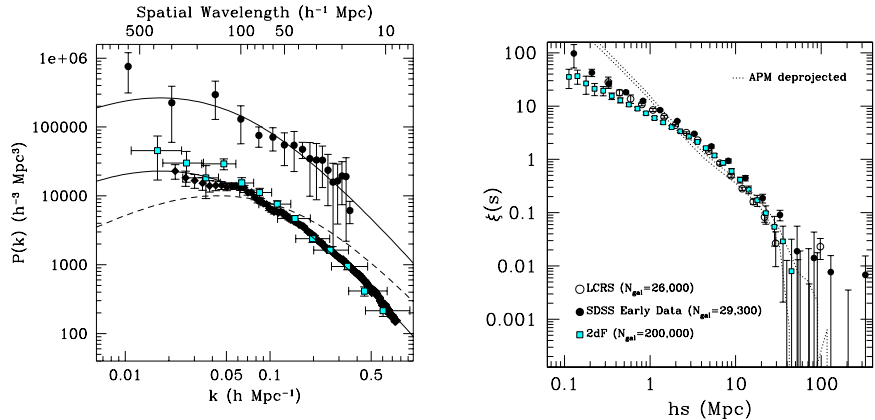} 
\caption{{\bf Left:} The power spectrum of 2dF galaxies and REFLEX
clusters.  {\it Filled diamonds}: estimate using 147,000 redshifts by
the 2dF team \cite{Percival2001}; {\it open squares}: Tegmark et
al. \cite{Tegmark_2dF} thorough analysis of the 100,000-redshift 2dF
public release (see Hamilton, this volume); {\it filled circles}:
REFLEX clusters in a $600\hmpc$ box \cite{reflex_pk}.  {\it Dashed
line}: Einstein-De Sitter CDM model; {\it lower solid line}:
Lambda-CDM `concordance' model (as defined in the first section); both
are normalized to match the amplitude of CMB fluctuations
\cite{Bunn_White}; {\it upper solid line}: same Lambda-CDM model, but
renormalized (``biased'') according to the mass distribution of REFLEX
clusters \cite{reflex_pk,BG2001}.  {\bf Right:} The two-point
correlation function from the 2dF \cite{Hawkins} and SDSS
\cite{Zehavi} galaxy data, compared to the previous Las Campanas
Redshift Survey \cite{LCRS_xi}. The dotted lines show instead a
correlation function in {\sl real space}, obtained through
deprojection from the APM angular galaxy catalogue \cite{Baugh96}
under two different assumptions about galaxy clustering evolution.}
\label{fig:pk}
\label{xi-surveys}
\end{figure}

The 2dF and REFLEX surveys have produced the best estimates to date of
the power spectrum of galaxies and X-ray clusters, respectively.
Fig.~\ref{fig:pk} (left panel) compares these data sets directly,
evidencing 
the remarkable similarity of the shape of $P(k)$ for these two classes.
This provides a direct confirmation of
the bias scenario, where clusters form at the rare high peaks of the
mass density distribution \cite{Kaiser84} and for this reason display
a stronger clustering amplitude.  
%
%
In the same figure I have also plotted the predictions for the mass power
spectrum of two models of the CDM family, computed as
described in \cite{Eisenstein_Hu}.  A model very close to what we
defined as the `concordance' model ($\Omega_M\simeq 0.3$,
$\Omega_\Lambda\simeq 0.7$, $h=0.7$)  provides in general an
excellent fit to the 2dF power spectrum, with a bias parameter
(i.e. normalization) close to unity\footnote{In fact, once we fix the
primordial spectrum $P_o$, in a pure CDM Universe the observed shape
depends only on $\Omega_M$, not on $\Omega_\Lambda$ (which on the other
hand influences the normalization).  In the literature, this is often
parameterized through a {\sl shape parameter} $\Gamma=\Omega_M\, h\,
f(\Omega_b)$, where $f(\Omega_b)\sim 1$ in case of negligible baryon
fraction.  The `concordance' model, therefore, has $\Gamma\simeq 0.2$.}.
The upper solid curve, on the
other hand, is the same model re-normalized as
$P_{clus}(k) = b_{clus}^2 P_{CDM}(k)$
where the cluster bias parameter $b_{clus}$ has been computed taking into
account the specific mass distribution function of the cluster sample, using a
relatively straightforward theory \cite{MW96,Sheth} (see
\cite{reflex_pk} for more details).  It is for these computations that
a well-understood mass selection function of our clustering tracers
is crucial.  The general result (an additional step with respect
to galaxies), is that our fiducial low-$\Omega_M$ CDM model
is capable to match very well {\bf both} the shape and amplitude of
the cluster $P(k)$ \cite{reflex_pk}.   The same shape agrees well also with
the power spectrum of the distribution of QSO's from the 2QZ survey, a
large redshift survey of colour-selected quasars carried out using the
same 2dF spectrograph at the Anglo-Australian Telescope \cite{2dF_QSO}.

As can be seen from fig.~\ref{fig:pk}, the low-$\Omega_M$ CDM model
predicts a maximum for $P(k)$ around $k=0.02 \kmpc$.  This turnover in
the power spectrum is an imprint of the horizon size at
the epoch of matter-radiation equality \cite{Paddy} and marks an
``homogeneity scale'', above which (smaller $k$'s) the variance drops 
below the white-noise behaviour.  In a pure fractal Universe, for
example, $P(k)$ would continue to rise when moving to smaller and smaller
$k$'s \cite{Guzzo97}. In fact, at least visually the data of
Fig.~\ref{fig:pk} do not really show a convincing indication for a
maximum.  In addition, on such extremely large scales ($\lambda > 500
\hmpc$), the effect of the survey geometry on the measured power can
be very significant, resulting in an effective survey {\sl window
function} in Fourier space which is convolved with the true underlying
spectrum (e.g. \cite{Tegmark_2dF}).  For highly asymmetric geometries,
the plane-wave approximation intrinsic in the Fourier decomposition
fails, and the convolution with the window function easily mimics a
turnover in a spectrum with whatever shape (e.g \cite{pk_esp}).  The
best solution in such cases is to resort to survey specific
eigenfunctions as those provided by the Karhunen-Loeve (KL) transform
\cite{Vogeley_Szalay}.  This technique has been intensively applied to
the REFLEX data \cite{KL_Peter}, confirming a shape corresponding to
$\Omega_M\simeq 0.2-0.4$ (corresponding to a turnover $k \simeq 0.02
\kmpc$).  
A more recent KL analysis using the joint constraints from
both the REFLEX mass function and power spectrum, provides a much
stronger result, $\Omega_M=0.34\pm 0.03$, together with a power spectrum
normalization $\sigma_8=0.71\pm 0.04 $ (expressed in the
conventional form as the variance within $8 \hmpc$-radius spheres)
\cite{KL_2}.  This normalization agrees very well with the
completely independent value obtained from a joint analysis of the 2dF
survey and CMB data\cite{Lahav02}.  

The SDSS will be the best data set to study the details of the power
spectrum around the peak scale. Preliminary results based on a small
subset of the survey \cite{Szalay_KL_SDSS,Dodelson_SDSS_pk} indicate a
best-fitting CDM spectral shape $\Gamma=0.14^{+0.11}_{-0.06}$,
i.e. virtually the same as measured by 2dF and REFLEX.  Again, we see
an impressive convergence of independent observations towards the same
low-$\Omega_M$ CDM model.  The SDSS luminous red-galaxy sample, in
particular, will be unique to investigate the
presence of {\sl baryonic features} in the power spectrum, produced by
oscillations in the baryonic matter component within the
last-scattering surface \cite{Eisenstein_Hu,Miller_peaks}.  These
features are expected to be of very low amplitude
unless the baryon density is much higher than currently established
and thus require a very high signal-to-noise and frequency resolution
in $P(k)$.  Potential wiggles seen in the 2dF power spectrum, in fact,
are shown to be an 
artifact of the survey window function and the strong bin-to-bin
correlation \cite{Percival2001,Tegmark_2dF}.

\subsection{The Two-Point Correlation Function}

%
%
%
In fact, the simplest statistics one can compute from the data and
also that for which the selection function is more directly
corrigible, is not the power spectrum, but rather its Fourier
transform, the {\sl two-point correlation function} $\xi(r)$, which
measures the excess probability over random to find a galaxy at a
separation $r$ from a given one
\cite{Peebles80}. Fig.~\ref{xi-surveys} (right panel) shows the correlation function
measured in {\sl redshift space}, $\xi(s)$ (see next section), from
the 2dF and SDSS current data sets \cite{Hawkins,Zehavi}, compared to
the LCRS \cite{LCRS_xi}.  Also shown (dotted lines) is the real-space
$\xi(r)$ reconstructed from the APM angular survey \cite{Baugh96}.
The shape
of $\xi(s)$ is roughly a power law  $\sim (s/s_o)^{-\gamma}$
between 0.1 and $30\hmpc$, with a {\sl correlation length} $s_o\simeq
8\hmpc$.  The overall difference with the $\xi(r)$ from the
APM survey (which is in real space, being based on a deprojection of
angular clustering), is due to redshift-space effects, that I
will address in detail in the next section.   Note how $\xi(s)$ maintains a
low-amplitude, positive value out to separations 
of more than $50\hmpc$, with the 2dF and SDSS data possibly implying a 
zero-crossing scale approaching $100\hmpc$.  This comparison shows
explicitly why large-size galaxy surveys are so important, given the
weakness of the clustering signal at such large separations\footnote{There
is quite a bit of confusion in technical papers on the term ``scale''
when comparing results from power spectra and correlation function
analyses.  A practical ``rule of thumb'' which works about right with
well-behaved spectra is that a scale $k$ in the power spectrum,
corresponding to a spatial wavelength $\lambda=2\pi/k$, relates
approximately to  $r\sim \lambda/4$ in $\xi(r)$.}.

\subsection{Velocity Distortions in the Redshift-Space Pattern}

The separation $s$ between two galaxies computed using their observed
redshifts is not a true distance: the red-shift observed in the galaxy
spectrum is in fact the quantity $cz=cz_{\rm
cosmological}+v_{\rm pec_{||}}$, where $v_{\rm pec_{||}}$ is the
component of the 
galaxy peculiar velocity along the line of sight.   This contribution
is typically of the order of $100\; {\rm km\,s^{-1}}$ for galaxies in the general
field, but can rise above $1000\; {\rm km\,s^{-1}}$ within high-density regions
as rich clusters of galaxies.
Fig.~\ref{xi-surveys} shows explicitly the consequence of such {\sl
redshift-space distortion} for the correlation function: $\xi(s)$ (all
points) is {\sl flatter} than its real-space counterpart (dotted
lines).  This is the result of two  
concurrent effects: on small scales ($r\mincir 2\hmpc$), clustering
is suppressed by high 
velocities in clusters of galaxies, that spread close pairs along the
line of sight producing what in redshift maps are sometimes called
``Fingers of God''.  
Some of these are perhaps recognisable in
Fig.~\ref{2dF_cones} as thin radial structures.
On the other hand, large-scale coherent streaming flows of galaxies
towards high-density structures enhance the apparent contrast of
these, when seen perpendicularly to the line of sight. This effect, on
the contrary, amplifies $\xi(s)$ above $\sim 3 - 5 \hmpc$.

%
\begin{figure}
\centering
\epsfxsize=12cm 
\epsfbox{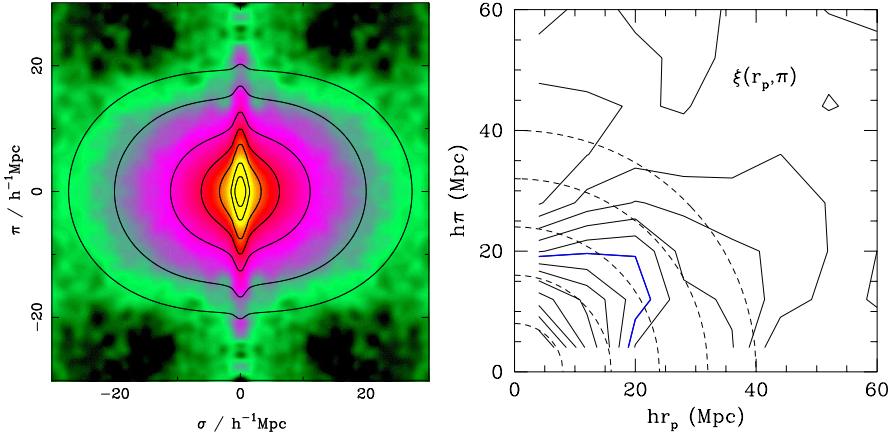} 
\caption{{\bf Left:} The bi-dimensional correlation function $\xi(r_p,\pi)$ from the 2dF
redshift survey (with $r_p$ named here $\sigma$) .  The 
large-scale deviation from circular symmetry is a measure of the level
of infall of galaxies onto superclusters, proportional to
$\beta=\Omega^{0.6}/b \simeq 0.43 $ \cite{2dF_xipz}.  {\bf Right:} the
same, but for the REFLEX survey of X-ray clusters of galaxies
(plotting only the first quadrant). Note also here the compression of
the contours along the redshift ($\pi$) direction, implying
significant streaming velocities of clusters 
towards high-density regions.  Note also the lack of any stretching at
very small $r_p$'s (there are no ``Fingers of God'' made by
clusters!) \cite{reflex_xi,Guzzo_csipz_REFLEX}.
}
\label{2dFbutterfly}
\label{fig:reflex_csipz}
\end{figure}
Such peculiar velocity contribution can be disentangled by
computing the two-dimensional correlation function $\xi(r_p,\pi)$,
where the separation vector $s$ between a pair of galaxies is decomposed
into two components, $\pi$ and $r_p$, parallel and perpendicular to
the line of sight respectively (see \cite{Hamilton} for details).
%
%
The result is a bidimensional map, whose iso-correlation contours look as
in Fig.~\ref{2dFbutterfly}, where the 2dF $\xi(r_p,\pi)$ is plotted \cite{2dF_xipz}. 

Redshift-space distortions contain important information as galaxy
motions are a  
direct dynamical probe of the mass distribution
\cite{Kaiser_dist}.  Non-linear distortions are
a measure of the ``temperature'' of the galaxy soup on small scales,
and they are in principle related to $\Omega_M$ through a {\sl Cosmic
Virial Theorem} \cite{Peebles80}, which however has been shown to be
difficult to apply in practice to real data \cite{Fisher94b}.  
Linear distortions produced by infall provide a way to measure the parameter
$\beta=\Omega_M^{0.6}/b$, i.e. essentially the mass density of the
Universe modulo the bias parameter.  As thoroughly explained in the excellent 
review by Andrew Hamilton \cite{Hamilton}, this can be achieved by
measuring the oblate compression of the contours of $\xi(r_p,\pi)$
along $\pi$.  One way to do this is to expand $\xi(r_p,\pi)$ in
spherical harmonics.  In linear perturbation theory, only the monopole
$\xi_0(s)$, quadrupole $\xi_2(s)$ and hexadecapole $\xi_4(s)$ are
non-zero, and it has been shown  \cite{Hamilton} that $\beta$ can in
principle be derived directly through  
%
%
a combination of these quantities. 
In practice, linear and non-linear effects are interlaced out to
fairly large scales ($\sim 20$ h$^{-1}$ Mpc), and require a careful
modeling also for a survey as large as the 2dF. This
has been done\footnote{A more careful analysis
applied to the 100,000 redshift public release \cite{Tegmark_2dF}, measures
essentially the same value of $\beta$, but with a 1-$\sigma$ error of
$\pm 0.16$, as discussed by Hamilton in his contribuion to this
volume.} 
\cite{2dF_xipz}, showing that the
quadrupole-to-monopole ratio of the map of Fig.~\ref{2dFbutterfly}
requires a Universe with $\beta=0.43\pm 0.07$.  If 2dF galaxies are
unbiased ($b\simeq 1$), this leads to $\Omega_M\simeq 0.25$. 

Clusters of galaxies clearly also partake in the overall motion of
masses produced by cosmological inhomogeneities.  Line-of-sight
spurious effects (as projections in optically-selected cluster
catalogues \cite{Collins95} or large redshift errors) and
limited statistics, prevented so far the detection of true velocity
anisotropies in cluster $\xi(r_p,\pi)$ maps.  Fig.~\ref{fig:reflex_csipz} plots
$\xi(r_p,\pi)$ for the REFLEX survey, which shows evidence for the 
compression of the contours along the line of sight, of the kind
expected by the linear infall of clusters towards superstructures
\cite{Guzzo_csipz_REFLEX}.

%
%

\section{Summary}

We are definitely in a golden age for observational cosmology and in
particular for the study of large-scale structure. We never had such a
wealth of diverse data at our disposal, through which we are pinning
down the values of cosmological parameters to a high accuracy
(e.g. \cite{CMB}).
The observational facts we have reviewed here
contribute to further reinforce the remarkable convergence among
different observables (CMB, large-scale structure, distant
Supernovae, cluster evolution, to mention a few) towards 
a model with
flat geometry ($\Omega_{total}=1$) provided by the
combination of a dominating Cold Dark Matter component
($\Omega_{M} = \Omega_{CDM}+\Omega_{baryon}\simeq 0.3$, with
$\Omega_{baryon}\simeq 0.04$) and a Dark
Energy of unknown nature (the cosmological constant,
$\Omega_{\Lambda}\simeq 0.7$ ).  

Still, we cannot avoid to note that such wonderful ``standard''
cosmological model is full of ``unseen'' ingredients, as a dark
matter nobody has detected so far (but see contribution by P. Belli
in this volume) and a dark
energy we have little idea where it could come from.  Seen from
outside, this might look as an almost epicyclic model and I believe
understanding its foundations provides one of the major challenges for
particle physics and cosmology in the next decade.

{\bf Acknowledgments.} I thank the organizers of the DARK2002 meeting
for inviting me to give this review.  I am grateful to all my
collaborators for all our common results discussed here, in particular
P. Schuecker, C. Collins and H. B\"ohringer.  I thank I. Zehavi and
E. Hawkins for providing their clustering results in electronic form.

%

\end{document}